\documentclass[aps,pra,reprint,superscriptaddress]{revtex4-2}
\usepackage{amsmath}
\usepackage{graphicx}
\usepackage{siunitx}

\begin{document}


\title{Hadamard Sensing Channel: Deterministic Artifact Suppression for Quantum Sensors}
\author{Weibin Ni}
\affiliation{Department of Physics, School of Science and Research Center for Industries of the Future, Westlake University, Hangzhou 310030, Zhejiang, China}
\affiliation{Institute of Natural Sciences, Westlake Institute for Advanced Study, Hangzhou 310024, Zhejiang, China}

\author{Lei Sun}
\email{sunlei@westlake.edu.cn}
\affiliation{Department of Physics, School of Science and Research Center for Industries of the Future, Westlake University, Hangzhou 310030, Zhejiang, China}
\affiliation{Institute of Natural Sciences, Westlake Institute for Advanced Study, Hangzhou 310024, Zhejiang, China}
\affiliation{Department of Chemistry, School of Science and Research Center for Industries of the Future, Westlake University, Hangzhou 310030, Zhejiang, China}

\begin{abstract}
Dynamical decoupling sequences are essential for nanoscale quantum sensing, but the finite duration of microwave pulses generates spurious responses. While phase randomization (PR) protocol can suppress these artifacts, they rely on probabilistic averaging. This introduces an inherent statistical variance that demands excessive sequence lengths and stringent hardware capabilities for true random phase generation. Here, we propose Hadamard sensing channel (HSC), a deterministic phase-design framework based on Hadamard matrices. HSC completely eliminates the statistical variance of PR by exactly and deterministically canceling spurious signals using a finite set of orthogonal phase patterns. Simulations confirm that HSC matches the ideal suppression of PR but exhibits superior robustness against control errors. HSC offers a mathematically exact and hardware-friendly solution for reliable high-resolution nanoscale nuclear magnetic resonance.
\end{abstract}

\maketitle

\newpage
\section{Introduction\label{intro}}
Dynamical decoupling (DD) stands as a cornerstone technique in quantum control, designed to protect quantum states from undesired system-environment interactions\cite{InformationProtectionRev,oneSecond,groupDD,Fmatrix,CHaDD}. By applying external control to qubits, DD substantially extends decoherence times and has unlocked profound applications in quantum sensing\cite{quantumSensing,fazhanShi}. Specifically, under DD control, nitrogen-vacancy (NV) centers in diamond can be engineered to act as narrow-band lock-in amplifiers capable of isolating specific AC magnetic fields\cite{FloquetSensing,DROID}, allowing the identification of nuclear spins at the nanoscale\cite{Sensing1,Sensing2,Sensing3,Sensing4}. Despite the widespread success of DD, the susceptibility of these sequences---even those designated as ``robust''---to control errors remains a pervasive yet largely overlooked challenge\cite{overestimation1,overestimation2,overestimation3,overestimation4,overestimation5}. These imperfections may lead to an overestimation of decoherence times\cite{CPC,HPC} and generate spurious harmonic responses that fundamentally compromise DD-based quantum sensing\cite{spuriousHarmonic,SRTheory}. Such artifacts severely spoil measurement fidelity; in extreme cases, they can falsely inflate apparent decoherence times by an order of magnitude or cause the direct misidentification of target signals, such as confusing genuine $^1\text{H}$ signals with $^{13}\text{C}$ artifacts in nanoscale nuclear magnetic resonance (NMR).   

One famous paradigm for suppressing these spurious resonances is phase randomization (PR), which relies on the application of random global phases to the fundamental pulse units\cite{PR}. Although PR demonstrates decent suppression performance in theory, its inherently probabilistic nature limits it to an approximate solution in practice. The probabilistic nature of PR makes the suppressed signal retain a residual statistical variance. This drives significant fluctuations in the detected signal, particularly when control errors are large or the sequence is short. Furthermore, generating a truly continuous spectrum of random phases imposes prohibitive hardware overheads. Specifically, generating arbitrary continuous phases require high-resolution arbitrary waveform generators (AWGs) and deep waveform memory.

Previously, we developed Hadamard phase cycling (HPC) for DD\cite{HPC}. HPC is a scalable quantum error mitigation method that leverages the abelian group structure of Hadamard matrices constructed by Sylvester's method\cite{Sylvester}. By systematically stacking four distinct matrix variations, HPC achieves near-quantitative error mitigation while maintaining a remarkably low, linear experimental complexity. This technique effectively filters out the adverse effects of control errors, such as coherence-population mixing, enabling the extraction of authentic decoherence times. In this work, we employ the Hadamard matrix to construct a sensing channel specifically designed to suppress the spurious responses induced by control errors. We refer to this deterministic and hardware-friendly framework for quantum sensing as the Hadamard Sensing Channel (HSC).

By systematically embedding discrete phase configurations derived from Hadamard matrices, the HSC architecture completely eliminates the statistical variance. Crucially, it significantly lowers the threshold for experimental realization. By strictly requiring a finite basis of standard orthogonal phases (e.g., $0, \pi/2, \pi, 3\pi/2$), it is natively compatible with modern digital quantum circuits and entirely obviates the need for the continuous phase spectrum demanded by PR. As our analysis reveals, this deterministic framework not only matches the theoretical suppression limits of ideal PR but exhibits superior robustness against both pulse angle and detuning errors. Ultimately, the HSC protocol provides a reliable, exact, and high-fidelity framework that can be universally applied to any DD sequence, paving the way for the robust quantum sensing.

\section{Spurious response\label{sr}}
In DD-based quantum sensing, a sensor qubit (e.g., an NV center in diamond) is subjected to a train of control pulses that modulates its coupling to the environment. We consider a typical Hamiltonian in the toggling frame with respect to the control pulses to introduce the spurious response effect:
\begin{equation}
\hat{H}(t)=F_z(t)\,\hat{S}_z \hat{E}(t) + \bigl[F_\perp(t)\,\hat{S}_- + \mathrm{H.c.}\bigr]\hat{E}(t),
\end{equation}
where $\hat{S}_z$ and $\hat{S}_-$ are the spin-1/2 operators of the sensor, and $\hat{E}(t)$ contains the target signal (e.g., an oscillating magnetic field or a nuclear spin) alongside environmental noise. The variables $F_z(t)$ and $F_\perp(t)$ are real-valued modulation functions determined by the pulse sequence. For ideal, instantaneous $\pi$ pulses, $F_\perp(t)= 0$ and $F_z(t)$ is a step function taking values of $\pm1$. Under these ideal conditions, the sensor resonantly couples only to signal components whose frequency matches that of $F_z(t)$. In practice, however, control pulses have finite durations. During each pulse, $F_z(t)$ deviates from $\pm1$ and, more critically, $F_\perp(t)$ acquires a nonzero value. This residual transverse modulation gives rise to spurious responses: a signal oscillating at a frequency not matched by $F_z(t)$ can still induce a detectable population change through the $F_\perp$-term, provided that its Fourier amplitude,
\begin{equation}
f_k^\perp = \frac{1}{T_{\mathrm{tot}}}\int_0^{T_{\mathrm{tot}}} F_\perp(t)\, e^{-i 2\pi k t/T_{\mathrm{tot}}}\,dt,
\end{equation}
is nonzero for some integer $k$, where $T_{\mathrm{tot}}$ is the total sensing time\cite{spuriousHarmonic,PR,SRTheory}. Such spurious peaks can mimic genuine signals, leading to the false identification of nuclear species (e.g., mistaking $^{13}$C for $^1$H) or erroneous estimations of sensitivity, thereby severely compromising the reliability of nanoscale NMR and magnetometry. 

To suppress these artifacts without sacrificing the desired resonance, one can modulate the phase of each pulse unit\cite{PR}. We begin our analysis by formalizing the unitary evolution of a single unit within a pulse sequence. Let $\hat{f}_k$ denote the free-evolution propagator between pulses, and let $\hat{P}(\phi)$ represent a pulse (e.g., a $\pi$ pulse) with phase $\phi$. For a sequence comprising $N_p$ pulses, the overall unitary operator is defined as:$$U_{\text{unit}}(\{\phi_j\}) = \hat{f}_{N_p+1}\, \hat{P}(\phi_{N_p})\, \hat{f}_{N_p} \cdots \hat{P}(\phi_2)\, \hat{f}_2\, \hat{P}(\phi_1)\, \hat{f}_1,$$where $\{\phi_j\}$ is the set of phases applied to the $N_p$ pulses. In the absence of phase modulation, all $\phi_j$ take a reference value of $0$ (Fig. 1(a)). A $\pi$ phase shift corresponds to $\phi_j \to \phi_j + \pi$, which flips the sign of the pulse operator. As discussed in Section \ref{intro}, adding a global randomized phase $\phi_R$ to each unit ($U_{\text{unit}}(\{\phi_j\}) \to U_{\text{unit}}(\{\phi_j + \phi_{R,j}\})$) suppresses these responses but introduces variance and demands statistical averaging and advanced hardware(Fig. 1(b)). We notice that the combined effect of a finite pulse duration and an AC magnetic field is physically analogous to a quantum state passing through a channel with detuning and rotation angle errors. Given that our previous work on Hadamard phase cycling for DD effectively addresses those errors, we propose adapting this deterministic framework into a sensing channel to entirely bypass the drawbacks of phase randomization (Fig. 1(c)). We refer to this sensing channel as Hadamard sensing channel (HSC).
\begin{figure}
	\centering
	\includegraphics[width=1.0\linewidth]{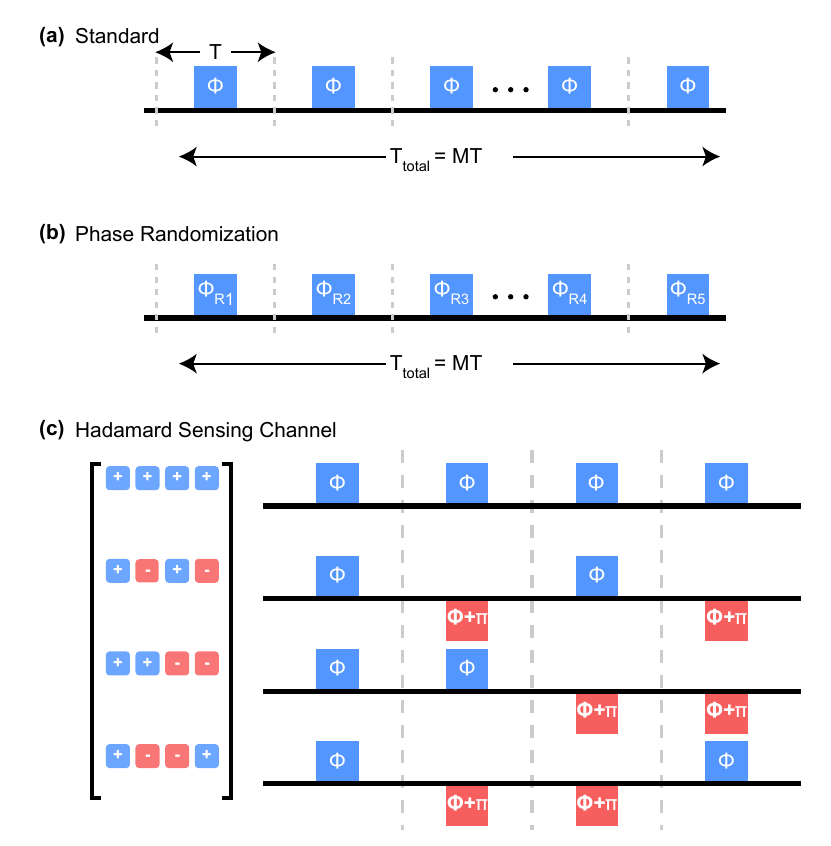}
	\caption{Schematic representations of various sequence modulation protocols. Rectangles represent pulse units. (a) The standard periodic DD sequence, consisting of repeated blocks of pulse units with a constant reference phase. (b) The phase-randomized periodic DD sequence, where a random phase offset is added to each block to suppress spurious responses, necessitating extensive statistical averaging. (c) The proposed HSC protocol applied to the periodic DD sequence. This panel illustrates the deterministic modulation across four distinct experimental runs, dictated by the rows of a 4-dimensional Sylvester-constructed Hadamard matrix ($H_4$). Blue rectangles represent pulse units with no additional phase modulation (a phase shift of $0$, corresponding to $+1$ matrix entries), whereas red rectangles denote pulses applied with a $\pi$ phase shift (corresponding to $-1$ matrix entries). The final error-mitigated sensing signal is extracted by deterministically averaging the experimental outcomes from all runs.}
	\label{fig1}
\end{figure}

\section{Hadamard sensing channel\label{hsc}}
Building upon this foundational concept, our current work introduces a streamlined adaptation. Crucially, unlike HPC which strictly requires the abelian group properties of Sylvester's construction, we discover that the suppression of spurious responses in quantum sensing only relies on the fundamental orthogonality of the Hadamard matrix. Consequently, any valid Hadamard matrix can be utilized.
\subsection{Introduction to the HSC}
We consider a pulse sequence consisting of $M$ identical pulse units, such as $M$ cycles of XY-8\cite{XY8}. In theory, the HSC protocol can expolit any Hadamard matrix to systematically assign global phases to pulse units across multiple experimental runs. For simplicity, we choose the Sylvester-constructed Hadamard matrix as an example. For $M=2^N$, it is generated recursively by:$$H_{2^N} = \begin{pmatrix} H_{2^{N-1}} & H_{2^{N-1}} \\ H_{2^{N-1}} & -H_{2^{N-1}} \end{pmatrix}, \qquad H_2 = \begin{pmatrix} 1 & 1 \\ 1 & -1 \end{pmatrix},$$with entries $s_{n, m}\in\{+1,-1\}$. Each row $n$ of $H_M$ defines a distinct phase configuration for the $M$ pulse units. For the $m$-th pulse unit in the $n$-th experimental run, we assign the phase:
$$\phi_m^{(n)} = \begin{cases} 0, & s_{n,m} = +1,\\ \pi, & s_{n,m} = -1. \end{cases}$$
Substituting these phases into our previously defined pulse unit equation and calculating the product of $M$ pulse units yields the total unitary evolution for a single experimental run:
\begin{equation}
U_n = \prod_{m=M}^{1} U_{\text{unit}}\bigl(\{\phi_j + \phi_m^{(n)}\}\bigr)
\end{equation}
The final HSC signal is obtained by averaging the expectation values of an observable $O$ over all $M$ unitaries:$$\langle O \rangle_{\rm HSC} = \frac{1}{M}\sum_{n=1}^{M} \operatorname{Tr}\Bigl[O\, U_n \rho U_n^{\dagger} \Bigr] = \operatorname{Tr}\Bigl[O\, \mathcal{E}_{\rm H}(\rho)\Bigr],$$where $\rho$ is the initial state, and the effective quantum channel is defined as:
\begin{equation}
\mathcal{E}_{\rm H}(\rho) = \sum_{n=1}^{M} p_n U_n \rho U_n^\dagger, \qquad p_n = \frac{1}{M},\quad \sum_{n=1}^{M} p_n = 1.
\end{equation}
This equation explicitly expresses the HSC operation as a convex combination of unitary channels. Since unitary channels are completely positive and trace-preserving (CPTP), and the set of CPTP maps is closed under convex combinations\cite{QIQC}, this formulation immediately certifies the HSC operation as a legitimate quantum channel. Our previous work indicates that this channel is particularly advantageous when pulse imperfections or free-evolution errors induce unwanted coherence–population transfers (spurious pathways) after a pulse unit\cite{HPC}. As a quantum state passes through the HSC, the vast majority of these spurious pathways acquire alternating phases across consecutive experimental runs, destructively averaging out. Precisely the same mechanism applies to the spurious response described in Section \ref{sr}: the residual transverse modulation $F_\perp(t)$ generates spurious coherence–population transfers, which are likewise deterministically suppressed by the HSC protocol.

\subsection{Theoretical proof}
While the mechanism intuitively suggests decent suppression, a rigorous mathematical proof confirms that HSC achieves the same benchmark suppression as PR in the weak coupling regime, but without the statistical variance. Consider $M$ periodic repetitions of a basic pulse unit with period $T$. The $k^{\mathrm{th}}$ Fourier amplitude of the modulation function is given by:
\begin{equation}
\begin{aligned}
	f^{\alpha}_{k} & = \frac{1}{MT}\int_{0}^{MT}F_{\alpha}(t)\exp\left(-i\frac{2\pi kt}{MT}\right)dt \\
	& = \frac{1}{MT}\sum_{m=1}^{M}\int_{(m-1)T}^{mT}F_{\alpha}(t)\exp\left(-i\frac{2\pi kt}{MT}\right)dt \\
	& = c_{k,M}\tilde{f}^{\alpha}_{k/M},
\end{aligned}
\end{equation}
where
\begin{equation}
	\tilde{f}^{\alpha}_{k/M} = \frac{1}{T}\int_{0}^{T}F_{\alpha}(t)\exp\left(-i\frac{2\pi k t}{M T}\right) dt,
\end{equation}
and
\begin{equation}
c_{k,M}=\frac{1}{M}\sum_{m=1}^{M}\exp\left(-i\frac{2\pi k(m-1)}{M}\right).
\end{equation}

In our protocol, a deterministic phase $s_{n,m}$ is applied to all pulses within the $m^{\mathrm{th}}$ basic pulse unit during the $n^{\mathrm{th}}$ experiment. This phase modulation—equivalent to a rotation about the $Z$-axis—transforms $F_\perp(t)$ to $F_\perp(t)s_{n,m}$ while leaving the $F_z(t)$ entirely unaffected, owing to its commutation with the $\sigma_z$ operator. It thus selectively modulates spurious responses while preserving the target signal. Consequently, the resulting transverse Fourier amplitude in the $n$-th experimental run becomes:
\begin{equation}
\begin{aligned}
	f^{\perp}_{n,k} & = \frac{1}{MT}\sum_{m=1}^{M}\int_{(m-1)T}^{mT}F_{\perp}(t)s_{n,m}\exp\left(-i\frac{2\pi kt}{MT}\right)dt \\
	& = H_{M}^{(n)}\tilde{f}^{\perp}_{k/M},
\end{aligned}
\end{equation}
where
\begin{equation}
H_M^{(n)} = \frac{1}{M} \sum_{m=1}^{M} s_{n,m} \exp\left(-i\frac{2\pi k(m-1)}{M}\right).
\end{equation}

For the $n^{\mathrm{th}}$ experiment, the intensity of the spurious responses is proportional to\cite{PR,SRTheory}:
$$\sin^2\!\left(\frac{1}{2}A_\perp\left\vert{}f_{n,k}^\perp\right\vert{}MT\right)\cos^2(\phi_k^\perp),$$
where $A_\perp$ denotes the perpendicular hyperfine coupling strength and $\phi_k^\perp$ is the complex phase of $f_k^\perp$. In the weak-coupling regime, this expression can be approximated as:
\begin{equation}
\left(\frac{1}{2}A_\perp\left\vert{}f_{n,k}^\perp\right\vert{}MT\right)^2\cos^2(\phi_k^\perp).
\end{equation}
Accordingly, the scaling of the spurious amplitude squared is:$$\left\vert{}f_{n,k}^\perp\right\vert{}^2 = \left\vert{}H_{M}^{(n)}\tilde{f}^{\perp}_{k/M}\right\vert{}^2 = \left\vert{}H_{M}^{(n)}\right\vert{}^2 \left\vert{}\tilde{f}^{\perp}_{k/M}\right\vert{}^2.$$

In the absence of phase cycling, $f_k^\perp=\tilde{f}^{\perp}_{k/M}$ when $k/M$ is an integer, and $f_k^\perp=0$ otherwise. Therefore, the suppression performance of the deterministic phase cycling is purely quantified by the average of $\left\vert{}H_M^{(n)}\right\vert{}^2$ over all experiments:
$$
	\frac{1}{M}\sum_{n=1}^{M} \left |H_M^{(n)}\right |^2
	=
	\frac{1}{M}\frac{1}{M^2}
	\sum_{n=1}^{M}
	\sum_{m,l=1}^{M}
	s_{n,m}s_{n,l}
	e^{-i2\pi k(m-l)/M}
$$
Exchanging the order of summation yields:
$$
	\frac{1}{M}\frac{1}{M^2}
	\sum_{m=1}^{M}
	\sum_{l=1}^{M}
	e^{-i2\pi k(m-l)/M}
	\sum_{n=1}^{M}
	s_{n,m}s_{n,l}.
$$
Exploiting the orthogonality of the Hadamard matrix, where:
\begin{equation}
	\sum_{n=1}^{M} s_{n,m} s_{n,l} =
	\begin{cases}
		M, & m = l, \\
		0, & m \neq l,
	\end{cases}
\end{equation}
we finally obtain:

\begin{equation}
\begin{aligned} 	
	\frac{1}{M} \sum_{n=1}^{M} \left\vert{}H_M^{(n)}\right\vert{}^2  	&= \frac{1}{M}\frac{1}{M^2} \sum_{m=1}^{M} e^{-i2\pi k(m-m)/M}M \\ 	&= \frac{1}{M}. 
\end{aligned}
\end{equation}
This mathematical result definitively shows that all spurious responses, independent of the Fourier index $k$, are consistently suppressed by a factor of $M$, the number of experimental runs. While this matches the suppression factor achieved by PR, HSC guarantees this suppression deterministically across the defined sequence, avoiding the statistical variance and hardware overhead inherent to random phase configurations.

\section{Simulation results\label{simulation}}
To validate the capability of the HSC protocol to suppress spurious responses, we first considered the simulation of a classical AC magnetic field. This sensing scenario can be equivalently modeled by a qubit coupled to a single nuclear spin, where the nuclear Larmor frequency $\omega_0/(2\pi)$ mimics the AC field frequency. The Hamiltonian for this model system is\cite{spuriousHarmonic}:
\begin{equation}
	\begin{aligned}
	\hat{H} = \underbrace{\Delta_z\hat{S}_z}_{\text{detuning}} &+  \underbrace{A_\perp\hat{S}_z\hat{I}_x}_{\text{AC signal}} + \omega_0 \hat{I}_z \\ &+ \underbrace{\omega_1\bigl[x(t)\hat{S}_x + y(t)\hat{S}_y\bigr]}_{\text{control}}.
	\end{aligned}
\end{equation}
In our simulations, the electronic sensor is initialized in the coherent state $\vert{+x}\rangle = (\vert{}0\rangle + \vert{}1\rangle)/\sqrt{2}$. The signal is quantified by the transition probability to the orthogonal $\vert{}-x\rangle$ state, which manifests as the distinct resonance peaks in the spectrum. We modeled a sensing sequence built from $M = 64$ XY-8 pulse units, equating to a total of 512 $\pi$ pulses. In a conventional, unmodulated XY-8 sequence, finite-pulse-duration effects induce a strong $F_\perp(t)$ modulation. This leads to the emergence of multiple spurious responses in the spectrum (clearly visible in Fig. 2(a)). By implementing the HSC protocol, all spurious artifacts are drastically suppressed, leaving only the desired target resonance and inherently ideal systemic peaks(Fig.2(b)). This demonstrates that HSC effectively averages out spurious responses without damping genuine signals. While the probabilistic PR protocol theoretically requires only a single experimental run, the inherent statistical variance dictates that multiple averaging runs are practically necessary to achieve a suppression level comparable to HSC. In this context, the multiple runs required by HSC do not constitute an extra overhead. Rather, by organizing these multiple runs deterministically through orthogonal phase patterns, HSC provides a structured averaging method that not only eliminates statistical fluctuations but also helps mitigate systematic pulse control errors in experiments.

\begin{figure}
	\centering
	\includegraphics[width=1.0\linewidth]{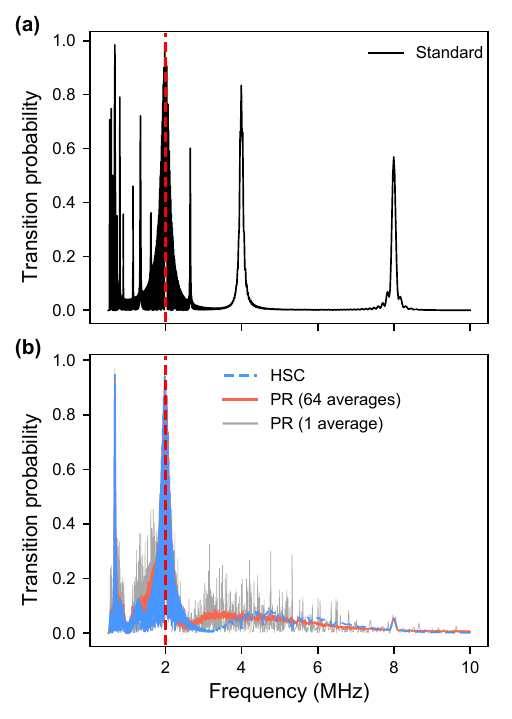}
	\caption{Simulated spectra for sensing a single nuclear spin using $64$ XY-8 cycles ($512$ pulses in total). Parameters: target frequency $f_{\text{ac}} = \omega_0/(2\pi) = 2$ MHz (highlight by red dashed line), coupling strength $A_\perp = 2\pi \times 200$ kHz, Rabi frequency $\omega_1/(2\pi) = 20$ MHz, and detuning $\Delta_z = 1$ MHz. (a) Standard XY-8 sequence without phase modulation, showing significant spurious responses. (b) Comparison of spectra between PR and the HSC. The HSC protocol inherently requires multiple sequence variations ($M = 64$). A single PR realization ($N_{\text{avg}} = 1$) exhibits significant statistical variance and averaging PR over 64 runs ($N_{\text{avg}} = 64$) effectively suppresses this variance.}
	\label{fig2}
\end{figure}

We next challenged the protocol with a more complex multi-nuclear scenario: a qubit hosted in an NV center in diamond coupled to two nuclear spins, $^1$H and $^{13}$C, where the explicit objective is to sense the $^1$H spin. The system Hamiltonian is written as\cite{PR}:
\begin{equation}
\begin{aligned}  	
	\hat{H} &= \underbrace{\Delta_z\hat{S}_z}_{\text{detuning}} + \underbrace{\omega_1\bigl[x(t)\hat{S}_x + y(t)\hat{S}_y\bigr]}_{\text{control}} \\ 	&\quad+ \underbrace{A_{\perp,1}\hat{S}_z\hat{I}_{x,1} + A_{\parallel,1}\hat{S}_z\hat{I}_{z,1} + \omega_{0,1}\hat{I}_{z,1}}_{^1\text{H signal}} \\  	&\quad+ \underbrace{A_{\perp,2}\hat{S}_z\hat{I}_{x,2} + A_{\parallel,2}\hat{S}_z\hat{I}_{z,2} + \omega_{0,2}\hat{I}_{z,2}}_{^{13}\text{C signal}}.
\end{aligned}
\end{equation}
Adopting the same $\vert{}+x\rangle$ initial state, the sensing signal is evaluated via the sensor's remaining coherence, which is directly given by the expectation value $\langle 2\hat{S}_x \rangle$.

 Here, $\omega_{0,1}/(2\pi) \approx 1.92$ MHz and $\omega_{0,2}/(2\pi) \approx 0.48$ MHz denote the Larmor frequencies of $^1$H and $^{13}$C, respectively, while $A_{\perp,k}$ represents their corresponding perpendicular hyperfine couplings. The primary challenge in this system is that the $^{13}$C spin generates a spurious response at precisely four times its Larmor frequency ($4 \times 0.48 \approx 1.92$ MHz). This artifact overlaps almost perfectly with the genuine $^1$H signal. In a standard XY-8 sequence, this creates ambiguous, overlapping peaks in the spectrum, rendering it practically impossible to distinguish the target $^1$H signal from the $4\times^{13}$C artifact. For our simulations, we used a train of $M=256$ XY-8 pulse units (2048 $\pi$ pulses). Assuming an ideal zero background detuning, we intentionally injected realistic static detuning ($\Delta_z$) and rotation amplitude errors, both expressed as a percentage of the Rabi frequency.

\begin{figure}
	\centering
	\includegraphics[width=1.0\linewidth]{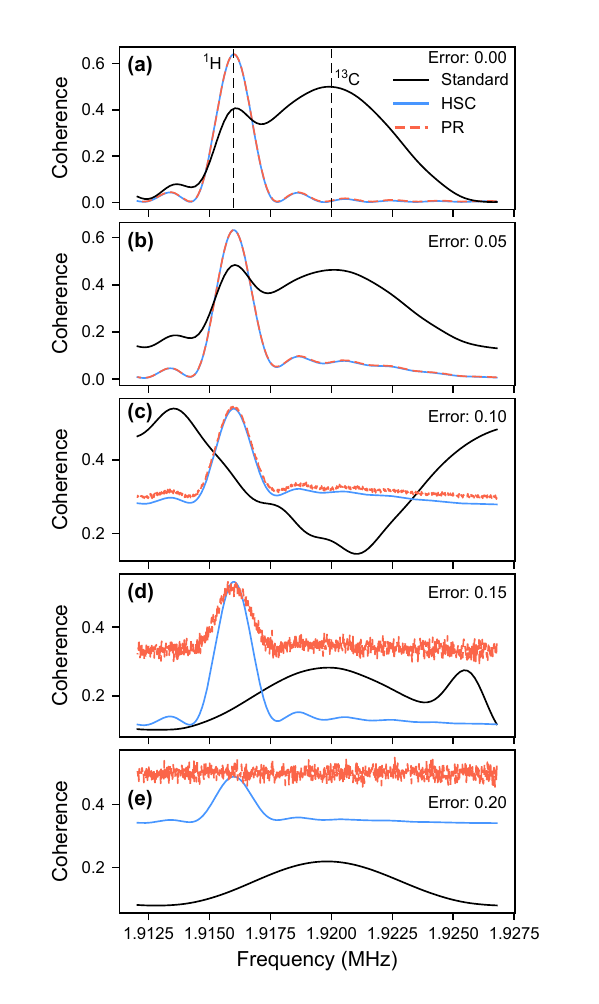}
	\caption{Simulated dual-species nuclear spin ($^1$H and $^{13}$C) sensing spectra using $M=256$ XY-8 cycles (yielding 2048 total $\pi$ pulses). The external magnetic field is $B = 45$ mT, corresponding to Larmor frequencies $\omega_{0,1}/(2\pi) = 1.92$ MHz ($^1$H) and $\omega_{0,2}/(2\pi) = 0.48$ MHz ($^{13}$C). The hyperfine coupling components $(A_{\perp}, A_{\parallel})$ are $2\pi \times (2, 1)$ kHz and $2\pi \times (20, 10)$ kHz for $^1$H and $^{13}$C, respectively. The nominal $\pi$-pulse duration is set to $200$ ns ($\omega_1/(2\pi) = 2.5$ MHz). Panels (a)–(e) compare the standard XY-8 sequence, PR, and HSC under increasing pulse error rates—comprising equal contributions from pulse amplitude and detuning errors, expressed as percentages of the nominal Rabi frequency $\omega_1$. The PR data is averaged over $N_{\text{avg}} = 256$ random realizations.}
	\label{fig3}
\end{figure}

Figure 3 compares the spectral outputs obtained via no phase modulation, the probabilistic PR protocol, and the deterministic HSC protocol across different error regimes. In the small-error regime (Figs. 3(a) and 3(b)), the standard protocol exhibits significant peak distortion because the target $^1$H response overlaps with the spurious response of the $^{13}$C spin. Both HSC and PR effectively suppress this spurious resonance with comparable performance. However, as control imperfections increase (Figs. 3(c)–(e)), the standard sequence completely fails to resolve the signals, and the signal contrast under PR degrades rapidly. This degradation occurs because PR relies on probabilistic averaging, meaning that these unwanted spurious responses are suppressed only up to a statistical variance. In a dense nuclear environment, this residual fluctuation compounds significantly, severely degrading the signal contrast and ultimately spoiling the measurement. HSC, by contrast, leverages the exact orthogonality of Hadamard matrices to deterministically suppress these spurious responses, completely bypassing the probabilistic noise floor. Consequently, HSC provides a demonstrably robust, deterministic solution for isolating genuine quantum signals in complex multi-nuclear environments, while maintaining remarkably low hardware demands. By requiring only four distinct control phases (e.g., $0, \pi/2, \pi, 3\pi/2$), HSC is natively compatible with standard IQ mixer hardware, completely eliminating the transient phase-switching errors and the need for high-resolution AWGs demanded by the continuous phase modulation of PR.

\section{Conclusion}
Spurious responses constitute a severe challenge in high-resolution quantum sensing. While phase randomization serves as a prominent mitigation strategy, its practical application is fundamentally hindered by two issues: a residual statistical variance that destabilizes measurements especially under large errors or short sequences, and the demanding hardware overhead required to generate true randomness.

The Hadamard sensing channel overcomes these limitations through a mathematically deterministic framework. Under ideal conditions, HSC exactly matches the theoretical suppression limits of PR. In the presence of realistic imperfections—specifically pulse angle errors and detuning, extending beyond mere finite pulse durations---HSC demonstrates significantly superior robustness.

Ultimately, our work extends beyond merely eliminating spurious resonances. It strongly reiterates the profound threat of control errors in dynamical decoupling, which not only artificially inflate apparent decoherence times but can also lead to complete sensing failures. By providing an exact, hardware-efficient solution, deterministic protocols like the HSC represent a pivotal step toward fundamentally resolving these vulnerabilities in quantum control and nanoscale magnetometry.

\begin{acknowledgments}
	This work was supported by the National Natural Science Foundation of China (grant No. 22273078) and Westlake University-YuYao Quantum Technology Joint Laboratory of Molecular Quantum Technology. W.N. and L.S. acknowledge Prof. Fazhan Shi, Zhijie Li and Guanyu Qu for helpful discussions. 
\end{acknowledgments}

\bibliography{REF}

@article{HPC, 
year = {2025}, 
title = {{Scalable quantum error mitigation with phase-cycled dynamical decoupling}}, 
author = {Ni, Weibin and Li, Zhijie and Qu, Guanyu and Equbal, Asif and Sun, Zhecheng and Dai, Jiale and Shi, Fazhan and Sun, Lei}, 
journal = {arXiv}, 
doi = {10.48550/arxiv.2511.12227}, 
}

@article{CPC, 
year = {2020}, 
title = {{Dynamical nuclear decoupling of electron spins in molecular graphenoid radicals and biradicals}}, 
author = {Lombardi, Federico and Myers, William K. and Ma, Ji and Liu, Junzhi and Feng, Xinliang and Bogani, Lapo}, 
journal = {Physical Review B}, 
issn = {2469-9950}, 
doi = {10.1103/physrevb.101.094406}, 
pages = {094406}, 
number = {9}, 
volume = {101}, 
}

@article{oneSecond, 
year = {2012}, 
title = {{Room-temperature quantum bit memory exceeding one second}}, 
author = {Maurer, P. C. and Kucsko, G. and Latta, C. and Jiang, L. and Yao, N. Y. and Bennett, S. D. and Pastawski, F. and Hunger, D. and Chisholm, N. and Markham, M. and Twitchen, D. J. and Cirac, J. I. and Lukin, M. D.}, 
journal = {Science}, 
issn = {0036-8075}, 
doi = {10.1126/science.1220513}, 
pmid = {22679092}, 
pages = {1283--1286}, 
number = {6086}, 
volume = {336}, 
}

@article{DROID, 
year = {2020}, 
title = {{Quantum metrology with strongly interacting spin systems}}, 
author = {Zhou, Hengyun and Choi, Joonhee and Choi, Soonwon and Landig, Renate and Douglas, Alexander M. and Isoya, Junichi and Jelezko, Fedor and Onoda, Shinobu and Sumiya, Hitoshi and Cappellaro, Paola and Knowles, Helena S. and Park, Hongkun and Lukin, Mikhail D.}, 
journal = {Physical Review X}, 
doi = {10.1103/physrevx.10.031003}, 
pages = {031003}, 
number = {3}, 
volume = {10}, 
}

@article{InformationProtectionRev, 
year = {2016}, 
title = {{Colloquium: Protecting quantum information against environmental noise}}, 
author = {Suter, Dieter and Álvarez, Gonzalo A.}, 
journal = {Reviews of Modern Physics}, 
issn = {0034-6861}, 
doi = {10.1103/revmodphys.88.041001}, 
pages = {041001}, 
number = {4}, 
volume = {88}, 
}

@article{fazhanShi, 
year = {2023}, 
title = {{Sub-nanotesla sensitivity at the nanoscale with a single spin}}, 
author = {Zhao, Zhiyuan and Ye, Xiangyu and Xu, Shaoyi and Yu, Pei and Yang, Zhiping and Kong, Xi and Wang, Ya and Xie, Tianyu and Shi, Fazhan and Du, Jiangfeng}, 
journal = {National Science Review}, 
issn = {2095-5138}, 
pmid = {37954192}, 
pmcid = {PMC10632795}, 
pages = {nwad100}, 
number = {12}, 
volume = {10}, 
}

@article{quantumSensing, 
year = {2017}, 
title = {{Quantum sensing}}, 
author = {Degen, C. L. and Reinhard, F. and Cappellaro, P.}, 
journal = {Reviews of Modern Physics}, 
issn = {0034-6861}, 
doi = {10.1103/revmodphys.89.035002}, 
pages = {035002}, 
number = {3}, 
volume = {89}, 
}

@article{spuriousHarmonic, 
year = {2015}, 
title = {{Spurious harmonic response of multipulse quantum sensing sequences}}, 
author = {Loretz, M. and Boss, J. M. and Rosskopf, T. and Mamin, H. J. and Rugar, D. and Degen, C. L.}, 
journal = {Physical Review X}, 
doi = {10.1103/physrevx.5.021009}, 
pages = {021009}, 
number = {2}, 
volume = {5}, 
}

@article{CHaDD, 
year = {2025}, 
title = {{Efficient chromatic-number-based multiqubit decoherence and crosstalk suppression}}, 
author = {Brown, Amy F. and Lidar, Daniel A.}, 
journal = {PRX Quantum}, 
doi = {10.1103/1d4l-73x6}, 
pages = {020354}, 
number = {2}, 
volume = {6}, 
}

@article{groupDD, 
year = {1999}, 
title = {{Dynamical decoupling of open Quantum systems}}, 
author = {Viola, Lorenza and Knill, Emanuel and Lloyd, Seth}, 
journal = {Physical Review Letters}, 
issn = {0031-9007}, 
doi = {10.1103/physrevlett.82.2417}, 
pages = {2417--2421}, 
number = {12}, 
volume = {82}, 
}

@article{Fmatrix, 
year = {2020}, 
title = {{Robust dynamic Hamiltonian engineering of many-body spin systems}}, 
author = {Choi, Joonhee and Zhou, Hengyun and Knowles, Helena S. and Landig, Renate and Choi, Soonwon and Lukin, Mikhail D.}, 
journal = {Physical Review X}, 
doi = {10.1103/physrevx.10.031002}, 
pages = {031002}, 
number = {3}, 
volume = {10}, 
}

@article{FloquetSensing, 
year = {2015}, 
title = {{Dynamical-decoupling-based quantum sensing: Floquet spectroscopy}}, 
author = {Lang, J. E. and Liu, R. B. and Monteiro, T. S.}, 
journal = {Physical Review X}, 
doi = {10.1103/physrevx.5.041016}, 
pages = {041016}, 
number = {4}, 
volume = {5}, 
}

@article{PR, 
year = {2019}, 
title = {{Randomization of pulse phases for unambiguous and robust quantum sensing}}, 
author = {Wang, Zhen-Yu and Lang, Jacob E. and Schmitt, Simon and Lang, Johannes and Casanova, Jorge and McGuinness, Liam and Monteiro, Tania S. and Jelezko, Fedor and Plenio, Martin B.}, 
journal = {Physical Review Letters}, 
issn = {0031-9007}, 
doi = {10.1103/physrevlett.122.200403}, 
pmid = {31172750}, 
pages = {200403}, 
number = {20}, 
volume = {122}, 
}

@article{SRTheory, 
year = {2017}, 
title = {{Enhanced resolution in nanoscale NMR via quantum sensing with pulses of finite duration}}, 
author = {Lang, J. E. and Casanova, J. and Wang, Z.-Y. and Plenio, M. B. and Monteiro, T. S.}, 
journal = {Physical Review Applied}, 
doi = {10.1103/physrevapplied.7.054009}, 
pages = {054009}, 
number = {5}, 
volume = {7}, 

}

@article{Sylvester, 
year = {1867}, 
title = {{LX. Thoughts on inverse orthogonal matrices, simultaneous signsuccessions, and tessellated pavements in two or more colours, with applications to Newton's rule, ornamental tile-work, and the theory of numbers}}, 
author = {Sylvester, J.J.}, 
journal = {The London, Edinburgh, and Dublin Philosophical Magazine and Journal of Science}, 
issn = {1941-5982}, 
doi = {10.1080/14786446708639914}, 
pages = {461--475}, 
number = {232}, 
volume = {34}, 
}

@article{Sensing1,
  title = {Sensing distant nuclear spins with a single electron spin},
  author = {Kolkowitz, Shimon and Unterreithmeier, Quirin P. and Bennett, Steven D. and Lukin, Mikhail D.},
  journal = {Phys. Rev. Lett.},
  volume = {109},
  issue = {13},
  pages = {137601},
  numpages = {5},
  year = {2012},
  month = {Sep},
  publisher = {American Physical Society},
}

@article{Sensing2,
author = {Simon Schmitt  and Tuvia Gefen  and Felix M. St{\"u}rner  and Thomas Unden  and Gerhard Wolff  and Christoph Müller  and Jochen Scheuer  and Boris Naydenov  and Matthew Markham  and Sebastien Pezzagna  and Jan Meijer  and Ilai Schwarz  and Martin Plenio  and Alex Retzker  and Liam P. McGuinness  and Fedor Jelezko },
title = {Submillihertz magnetic spectroscopy performed with a nanoscale quantum sensor},
journal = {Science},
volume = {356},
number = {6340},
pages = {832-837},
year = {2017}}

@article{Sensing3,
author = {T. Staudacher  and F. Shi  and S. Pezzagna  and J. Meijer  and J. Du  and C. A. Meriles  and F. Reinhard  and J. Wrachtrup },
title = {Nuclear magnetic resonance spectroscopy on a (5-Nanometer)$^3$ Sample Volume},
journal = {Science},
volume = {339},
number = {6119},
pages = {561-563},
year = {2013}}

@article{Sensing4,
author = {I. Lovchinsky  and A. O. Sushkov  and E. Urbach  and N. P. de Leon  and S. Choi  and K. De Greve  and R. Evans  and R. Gertner  and E. Bersin  and C. Müller  and L. McGuinness  and F. Jelezko  and R. L. Walsworth  and H. Park  and M. D. Lukin },
title = {Nuclear magnetic resonance detection and spectroscopy of single proteins using quantum logic},
journal = {Science},
volume = {351},
number = {6275},
pages = {836-841},
year = {2016},
doi = {10.1126/science.aad8022},
URL = {https://www.science.org/doi/abs/10.1126/science.aad8022}}

@article{overestimation1, 
year = {2013}, 
title = {{Solid-state electronic spin coherence time approaching one second}}, 
author = {Bar-Gill, N. and Pham, L.M. and Jarmola, A. and Budker, D. and Walsworth, R.L.}, 
journal = {Nature Communications}, 
doi = {10.1038/ncomms2771}, 
pmid = {23612284}, 
pages = {1743}, 
number = {1}, 
volume = {4}, 
}

@article{overestimation2, 
year = {2026}, 
title = {{Optically Addressable Molecular Spins at 2D Surfaces}}, 
author = {Zhou, Xuankai and Kong, Yan-Tung and Cheung, Cheuk Kit and Bian, Guodong and Moukaouine, Reda and Wong, King Cho and Sun, Yumeng and Ho, Cheng-I and Bushmakin, Vladislav and Gross, Nils and Yen, Chun-Chieh and Priessnitz, Tim and Lenger, Malik and Jayaram, Sreehari and Taniguchi, Takashi and Watanabe, Kenji and Pershin, Anton and Peng, Ruoming and Gali, {\'A}d{\'a}m and Smet, Jurgen and Wrachtrup, J{\"o}rg},
journal = {arXiv}, 
doi = {10.48550/arxiv.2601.19988}
}

@article{overestimation3, 
year = {2026}, 
title = {{A Surface-Scaffolded Molecular Qubit}}, 
author = {Zheng, Tian-Xing and Utama, M Iqbal Bakti and Gao, Xingyu and Kar, Moumita and Yu, Xiaofei and Kang, Sungsu and Cai, Hanyan and Ruan, Tengyang and Ovetsky, David and Zvi, Uri and Lao, Guanming and Wang, Yu-Xin and Raz, Omri and Chitransh, Sanskriti and Smith, Grant T and Weiss, Leah R and Czyz, Magdalena H and Yang, Shengsong and Fairhall, Alex J and Watanabe, Kenji and Taniguchi, Takashi and Awschalom, David D and Alivisatos, A Paul and Goldsmith, Randall H and Schatz, George C and Hersam, Mark C and Maurer, Peter C}, 
journal = {arXiv}, 
doi = {10.48550/arxiv.2601.19976}, 
}

@article{overestimation4, 
year = {2024}, 
title = {{Long-lived coherences in strongly interacting spin ensembles}}, 
author = {Schenken, William K. and Meynell, Simon A. and Machado, Francisco and Ye, Bingtian and McLellan, Claire A. and Joos, Maxime and Dobrovitski, V. V. and Yao, Norman Y. and Jayich, Ania C. Bleszynski}, 
journal = {Physical Review A}, 
issn = {2469-9926}, 
doi = {10.1103/physreva.110.032612}, 
pages = {032612}, 
number = {3}, 
volume = {110}, 
}

@article{overestimation5, 
year = {2025}, 
title = {{Quantum sensing with spin defects in boron nitride nanotubes}}, 
author = {Rizzato, Roberto and Hidalgo, Andrea Alberdi and Nie, Linyan and Blundo, Elena and von Grafenstein, Nick R. and Finley, Jonathan J. and Bucher, Dominik B.}, 
journal = {Nature Communications}, 
doi = {10.1038/s41467-025-67538-2}, 
pmid = {41419738}, 
pmcid = {PMC12722255}, 
pages = {11333}, 
number = {1}, 
volume = {16}, 
}

@article{XY8, 
year = {1990}, 
title = {{New, compensated Carr-Purcell sequences}}, 
author = {Gullion, Terry and Baker, David B and Conradi, Mark S}, 
journal = {Journal of Magnetic Resonance (1969)}, 
issn = {0022-2364}, 
doi = {10.1016/0022-2364(90)90331-3}, 
pages = {479--484}, 
number = {3}, 
volume = {89}, 
}

@book{QIQC,
  title={Quantum Computation and Quantum Information},
  author={Nielsen, Michael A and Chuang, Isaac L},
  year={2000},
  publisher={Cambridge University Press}
}
\end{document}